# Subsidence and current strain patterns on Tenerife Island (Canary Archipelago, Spain) derived from continuous GNSS time series (2008–2015)

A. Sánchez-Alzola [a],*, J. Martí [b], A. García-Yeguas [c], A.J. Gil [d]

[a] Departamento de Estadística e Investigación Operativa, Universidad de Cádiz, Spain
[b] Instituto de Ciencias de la Tierra "Jaume Almera" CSIC, Barcelona, Spain
[c] Departamento de Física Aplicada, Universidad de Cádiz, Spain
[d] Departamento de Ingeniería Cartográfica, Geodésica y Fotogrametría, Universidad de Jaén, Spain

* Corresponding author.
  E-mail address: alberto.sanchez@uca.es (A. Sánchez-Alzola).

a b s t r a c t

In this paper we present the current crustal deformation model of Tenerife Island derived from daily CGPS time series processing (2008–2015). Our results include the position time series, a global velocity estimation and the current crustal deformation on the island in terms of strain tensors. We detect a measurable subsidence of 1.5–2 mm/yr. in the proximities of the Cañadas-Teide-Pico Viejo (CTPV) complex. These values are higher in the central part of the complex and could be explained by a lateral spreading of the elastic lithosphere combined with the effect of the drastic descent of the water table in the island experienced during recent decades. The results show that the Anaga massif is stable in both its horizontal and vertical components. The strain tensor analysis shows a 70 nstrain/yr. E-W compression in the central complex, perpendicular to the 2004 sismo-volcanic area, and 50 nstrain/yr. SW-NE extension towards the Northeast ridge. The residual velocity and strain patterns coincide with a decline in volcanic activity since the 2004 unrest.

# 1. Introduction

Tenerife Island (Canary Islands, Spain) has been the subject of detailed geological and geophysical research because of its volcanological evolution and geodynamic setting. After a century-long period of quiescence the central volcanic complex of the island (Teide Pico Viejo stratovolcanoes) reactivated in the spring of 2004 with an increase of seismicity with over 3000 recorded events, five of them felt by the population. This activity was generally explained by magma addition into its northwestern area (Tárraga et al., 2006; Gottsmann et al., 2006, 2008; Almendros et al., 2007; Martí et al., 2009). Due to its great interest, several studies have been developed after this reactivation, focusing on aero-magnetic surveying in order to study the structure and growth of the island (Blanco-Montenegro et al., 2011, García et al., 2007), gravity models to analyze the structure beneath the central volcanic complex (Gottsmann et al., 2008), seismological studies using scattering seismic attenuation and imaging magma storage (Ibañez et al., 2008; García-Yeguas et al., 2012; Lodge et al., 2012; De Barros et al., 2012; Prudencio et al., 2013, 2015), resistivity measurements (Coppo et al., 2008; Piña-Varas et al., 2014), InSAR and vertical deformation (Fernández et al., 2009; Tizzani et al., 2010), tiltmeter measurements (Eff-Darwich et al., 2008) and episodic DGPS deformation analysis (Berrocoso et al., 2010).

In particular, spatial geodetic techniques have been a reliable and valuable tool for determining velocity gradients and deformation patterns on the surface. Currently GPS data and continuous position time series analysis have been thoroughly used in ground deformation and monitoring of active volcanoes as a complement to traditional monitoring systems (Myer et al., 2008; Perlt et al., 2008; Shirzaei et al., 2013; Marques et al., 2013). Geodetic tools have enabled a better understanding of these active areas and facilitated precise measurements of crustal displacement and derived strains.

In this contribution we present a rigorous crustal deformation model of Tenerife Island derived from CGPS position time series using Precise Point Positioning (PPP). We also include a velocity field analysis and a strain rate array with the data available. The results were integrated with geological information and previous studies to improve the current geodynamical analysis of the Island. A detailed assessment using newly continuous GNSS stations is of great importance in improving our knowledge of the shallow sub-surface of Tenerife and the current situation after the 2004 reactivation event.

# 2. Geological settings

Tenerife Island, situated close to the NW coast of Africa in the Atlantic Ocean, is the major volcanic island of the Canary Islands archipelago (Spain) (Fig. 1). Due to its volcanic evolution, the geomorphology of Tenerife is quite heterogeneous. The presence of varied volcanic processes throughout its history have given quite a few examples of basaltic lava flows, strato-volcanoes, monogenetic cones, pyroclastic deposits and calderas (Las Cañadas). The most significant structure is the Las Cañadas-Teide-Pico Viejo (CTPV) complex which is located in the centre of the island. This formation rises nearly 4000 m above sea level and comprises a collapse caldera and a stratovolcano complex (Teide Pico Viejo) (Ablay and Martí, 2000). Its core is thought to be ultramafic cumulitic (Ablay and Kearey, 2000; Gottsmann et al., 2008), and it is responsible for gravity, magnetic and seismic velocity anomalies (Watts et al., 1997; Ablay and Kearey, 2000; Araña et al., 2000). Surrounding this central complex we observe two main ridges NW-SE and NE-SW (Martí et al., 1994) and a subsidiary monogenetic basaltic field at the southern sector of the island (Geyer and Martí, 2010). The volcanism of the Canary Archipelago has been affected by several regional fault systems. These faults are classified into two different families according to their origin: The features related to the opening of the Atlantic (NE-SW direction) and those influenced by the tectonics of the Atlas range in the African continent (NW-SE). These regional fault systems have an influence on the entire tectonic framework of Tenerife and determine the location of the eruptions on the whole island (Ablay and Martí, 2000; Geyer and Martí, 2010). During the Holocene volcanic activity, two kinds of volcanism occurred in Tenerife (Martí et al., 2008):

(1) a basaltic fissure volcanism, located mainly on the ridges and the south of the island and (2) phonolitic eruptions related to the CTPV complex. The record of historical eruptions includes the Sietefuentes, Fasnia and Arafo volcanoes (1704–1705) in the NE-SW ridge and Garachico (1706) and El Chinyero (1909) volcanoes in the NW-SE. The Chahorra eruption (1798) is located outside the ridge areas, on the western flank of the Pico Viejo stratovolcano (Romero, 1992). The active Teide-Pico Viejo twin stratovolcanoes started to grow inside the Las Cañadas caldera about 180 k ago (Martí et al., 1994) and have mainly produced lava flow and explosive eruptions of basaltic to phonoltic magmas. The products from these eruptions partially fill the caldera, and the adjacent Icod and La Orotava valleys, to the north, both originated by landslide processes related to the caldera formation (Martí et al., 1997).

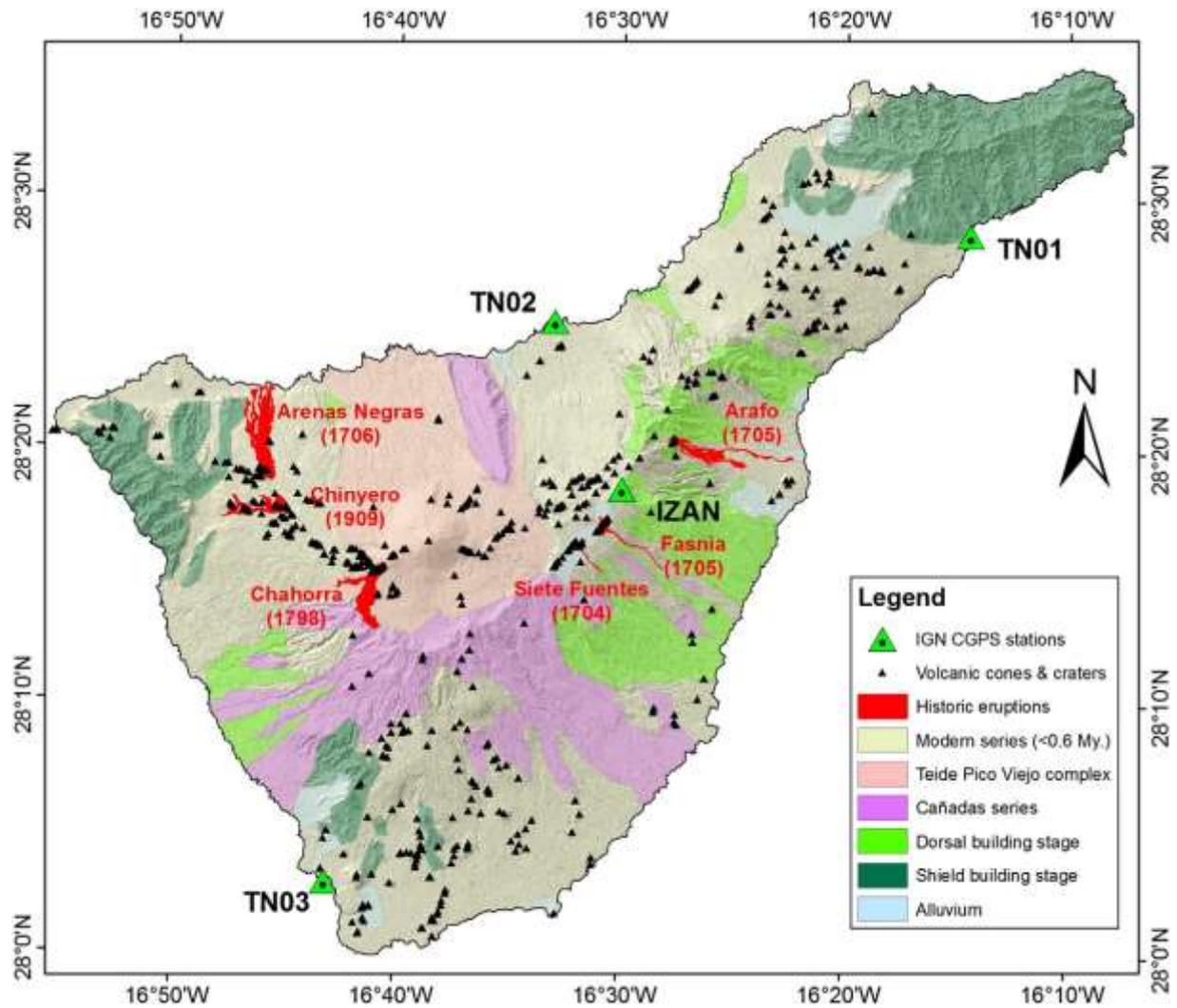

Fig. 1. Schematic geological map of Tenerife Island (Canary Archipelago) (based on Carracedo et al., 2007). The historical lava flows (in red), recent volcanic cones and craters (black dots) and the CGPS stations used in this study (green triangles) are indicated.

3. CGPS data and processing

3.1. CGPS stations

The ERGNSS network (Red de Estaciones de Referencia GNSS) is an active continuous geodetic network belonging to the public Instituto Geográfico Nacional of Spain. This network has a national coverage, is included in the International Terrestrial Reference Frame (ITRF) and is framed within the A-class EUROGEOGRAPHICS network. Its primary objective is to provide RINEX data and support users in geodetic, cartographic, and topographic studies.

For this study we have used the four CGPS ERGNSS stations currently available and distributed across Tenerife Island. These stations have been operated by the Spanish IGN since 2008 (Fig. 1): TN01, built over the Mareograph building of the north wharf in Santa Cruz de Tenerife; TN02 built on the roof of the fishermen's guild building of Puerto de la Cruz; TN03 situated on the Port Authority building of Los Cristianos (Arona); and IZAN, built on a concrete pillar near the Meteorological observatory of Izaña. All of these stations have LEICA GRX1200GGPRO receivers and LEIAT504GG leica choke ring antennas. They also have data available after 2008 and have not been subjected to changes of antenna in the time span 2008–2015. The IZAN station has also been included in the EUREFF Permanent Network (EPN) since 2009. TN01 is situated on the eastern side of the Dorsal rift near Anaga. This area is the transition between the basaltic massif to the rift and is characterized by few volcanic emissions. TN02 is located in the north sector of La Orotava valley. IZAN is located on the south-occidental side of the NE-SW dorsal rift over deposits of altered basaltic pyroclasts. Finally, TN03 is situated on the southern slope of the Cañadas edifice near the southern ignimbrite deposits of Tenerife. Fig. 2 shows the distribution of the CGPS ERGNSS stations across Tenerife Island and the location of the earthquakes during the 2004 unrest and the period studied, 2008–2015.

3.2. CGPS position time series

In order to process and generate the position time series of the CGPS stations we used version 6.2 of GIPSY-OASIS, developed by the Jet Propulsion Laboratory (https://gipsy-oasis.jpl.nasa.gov/). This program has a basic and robust module named gd2p.pl that works with single receiver data to generate a static Precise Point Positioning coordinate. For better determination of the time series, we also used 24 h 30-second sampling interval RINEX observation files accumulated over a seven year period (January 2008–march 2015). The software takes these observation files, applies certain basic flags joined to the module, and processes the data downloading all the files needed for the PPP processing.

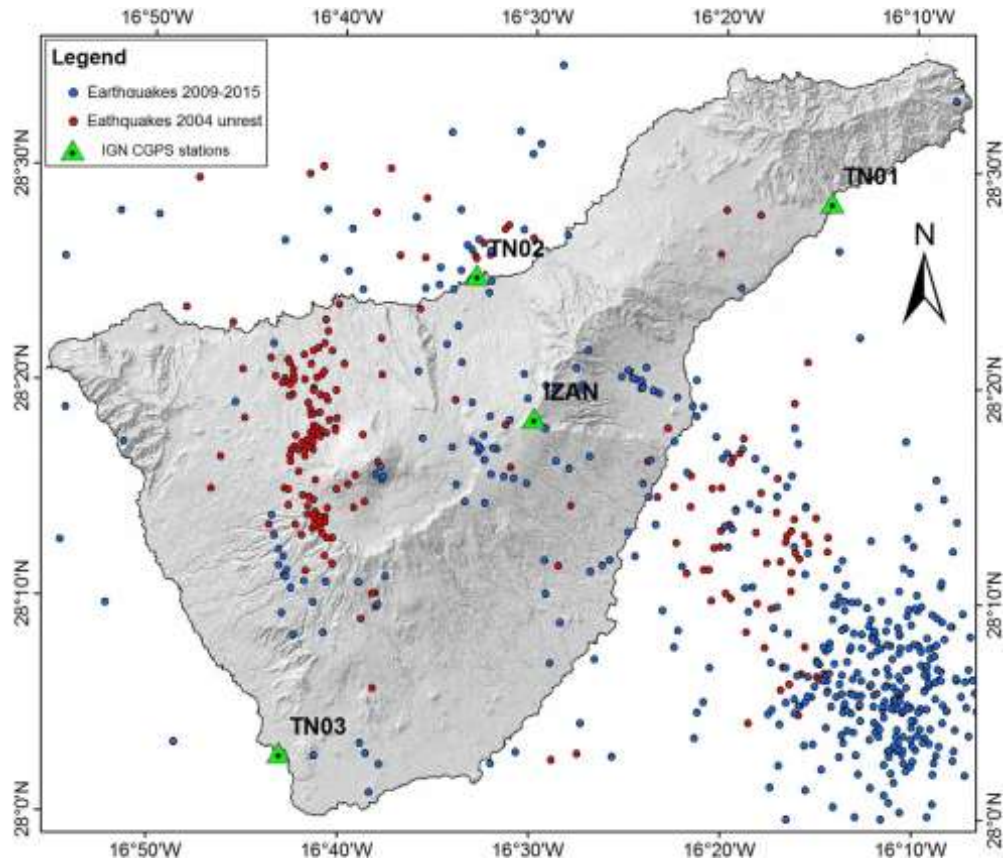

Fig. 2. Seismicity on Tenerife Island during 2004 volcanic crisis (red dots) the period of study 2008–2015 (blue dots), based on the IGN database (http://www.ign.es/ign/layoutIn/ sismoPrincipalTensorZonaAnio.do) and (Domínguez Cerdeña et al., 2011). ERGNSS CGPS stations (green triangles) are also indicated.

Concerning the methodology, a standard GIPSY-OASIS procedure with the gd2p.pl module was applied to all of the data sets to generate the position time series. First, GIPSY-OASIS downloaded the final ephemeris and pole products from the JPL FTP repository into a single IGS08 coordinate system. In order to fix the ambiguities with zero difference in the PPP processing, we used the strategy described in Bertiger et al. (2010) considering the wide lane phase bias file (wlpb) This method considers a least squares fit of sinusoids to data samples that boots long-periodic noise in gapped records (Press and Rybicki, 1989). The periodogram is obtained using a τ parameter that represents the time delay after which a pair of considered sinusoids is mutually orthogonal at sample time. The amplitude is computed as a function of angular frequencies ($\omega = 2\pi f$). available on the JPL. Furthermore, The GMF troposphere mapping function (Boehm et al., 2006) and ionosphere TEC values were also used. In addition, the FES2004 Ocean tide-loading model (Lyard et al., 2006) from the Onsala Space Observatory (http://holt.oso.chalmers.se/ loading/) and the WahrK1 terrestrial tide model (Wahr, 1985) were considered. We also modelled the hydrostatic component of the zenith tropospheric delay and estimated the wet component. The suitable

IGS08week.atx antenna calibration file from the JPL was applied in order to correct the Antenna Phase Centre. To avoid multipath, an elevation mask of 10° was set up as well. A 300-second sampling of the RINEX observations for the coordinate estimation was also used. Eventually, we did not consider any network adjustment, only selecting daily PPP solutions and applying no dependencies between stations.

$$\tan 2\omega = \frac{\sum_j \sin 2\omega t_j}{\sum_j \cos 2\omega t_j} \quad P_x(\omega)$$

$$= \frac{1}{2}\left(\frac{[\sum_j X_j \cos 2\omega(t_j-\tau)]^2}{\sum_j \cos 2\omega(t_j-\tau)} + \frac{[\sum_j X_j \sin 2\omega(t_j-\tau)]^2}{\sum_j \sin 2\omega(t_j-\tau)}\right)$$

To manage the time series, the computed position results were converted from their raw XYZ format (geocentric coordinates) to their horizontal and vertical components. Prior to periodicity and white noise estimation on the time series, certain outliers were also eliminated using a specific threshold based on the standard deviation of the adjusted model. The series generated were free of any offsets or jumps due to changes of antennas. Furthermore, the positions were computed in the same coordinate system (IGS08) due to the reprocessing and correct maintenance by the JPL of its products.

Although GNSS coordinate time series were obtained using daily RINEX files, some periods are incomplete and uneven in time. To avoid problems in the harmonic estimation we used a methodology based on the Lomb Scargle periodogram (Lomb, 1976; Scargle, 1982). To estimate the harmonics in this study we used the de-trended time series and the Lomb Scargle normalized periodogram. This technique was implemented in a self-programming module with Matlab software following the source code described in Press et al. (1992). Fig. 3 shows the frequency in cycles per year and the spectral power in cm$^2$/cpy. in horizontal components East (blue) and North (red) and vertical component Up (grey).

### 3.3. CGPS-derived velocity field and strain rates

We employ the CATS GPS coordinate time series analysis software (Williams, 2008) to estimate the time series parameter. This is a command-line orientated program which applies a maximum likelihood estimation (MLE) and fits a multi-parameter model to the time series software. In this study we considered annual and semiannual harmonics joined to the trend of the station that defines its velocity (Penna and Stewart, 2003; Teferle et al., 2007). This agrees with the main harmonics obtained with the previous Lomb Scargle periodicity analysis. In order to propagate the uncertainties of the absolute velocities we considered the contribution of the white noise and the flicker noise from the time series (Mao et al., 1999; Williams et al., 2004). We modelled the covariance matrix for flicker noise cross at a period of one year with daily sampling intervals and equal amplitude. The spectral indices were computed from periodogram and included in the error analysis. We also propagated the uncertainties obtained in the GPS processing into the accuracy of the velocity estimation.

Afterwards, we obtained the residual velocity field in order to highlight the differences between the absolute horizontal velocities and the local displacements. In this study we considered the ITRF2008 plate model (Altamimi et al., 2011). For this purpose, we used the transformation parameters of the ITRF2008 plate model, considering the Nubian plate as reference and the geographical coordinates of the stations as inputs. In order to perform a correct analysis we also included the decline of the phreatic level on the Island from 1925 to 1997 from Custodio et al. (2016). Fig. 4 shows the de-trended position time series from the CGPS stations (Horizontal and vertical components are in meters). Fig. 5 includes the residual velocity field without the plate effect obtained in both components and the decline of the phreatic level. Table 1 shows the values of the absolute and residual velocities and the uncertainties computed (all in mm/yr.).

We also calculated the strain tensors based on the absolute velocity field we previously computed, using the GRID_STRAIN (Teza et al., 2008) Matlab software module. Fig. 6 shows the principal axes of the strain tensors computed, showing only significant grid points. The blue colour indicates extension and the red one compression. We also computed the strain errors using the same software. Error values fluctuate between 0.2 and 96.5 nstrain/yr. over a maximum strain of 134.7 nstrain/yr.

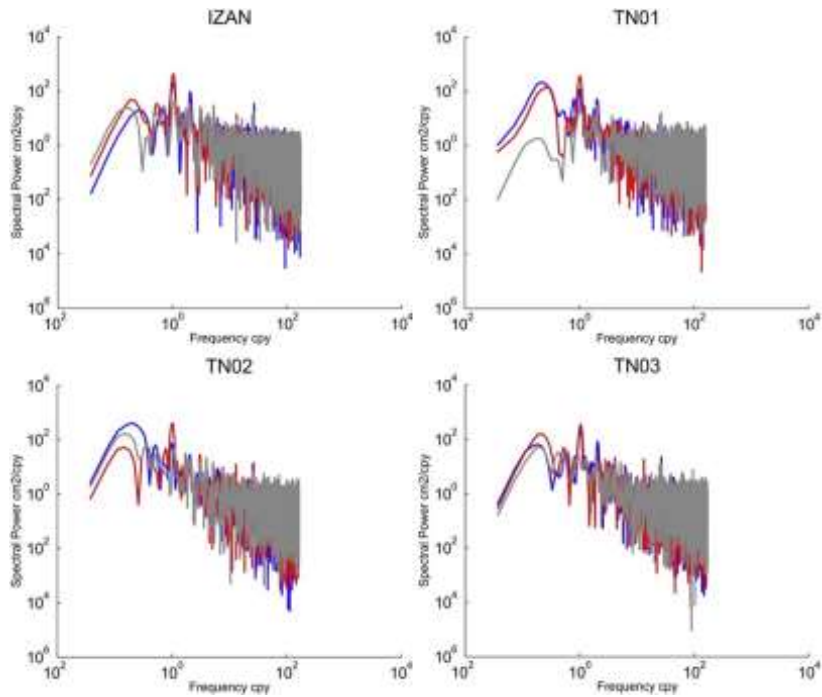

Fig. 3. Frequency in cycles per year and spectral power in cm$^2$/cpy of the CGPS stations. Components East (blue), North (red) and Up (grey). We observe a prominence of annual (1) and semiannual (0.5) periodicities in all the components.

Results and discussion

*3.4.* Time series and velocity field

The methodology described in this paper was applied to all the GNSS observations obtained during the study period (2008–2015). The strains and velocity field generated are the best possible model considering the data and the CGPS stations available during this 7 year observational period. The rigorous study of periodicities with the Lomb-Scargle normalized periodogram shows a predominance of annual and semiannual harmonics in both horizontal and vertical components in the time series (Fig. 3). These results agree with other GNSS deformation studies using CGPS stations (Sánchez-Alzola et al., 2014; Khelifa et al., 2013) and validate the stability of the model. The location of the sites over steady concrete monuments and the use of choke-ring antennas ensure the best quality of signals and the subsequent time series analysis.

The small deformations obtained are usual in other volcanic systems (Mattioli et al., 1998; Palano et al., 2008; Wadge et al., 2006). The horizontal velocity field reveals that stations situated near the CTPV complex (TN02, TN03 and IZAN) have residuals of 0.4–1.0 ± 0.2 mm/yr. with respect to Nubia plate ITRF2008 velocities (Table 1). TN01 is the most stable station with a velocity close to the tectonic environment (residual velocity of 0.21 ± 0.16 mm/yr.). We consider that the velocity residuals close to the CTPV complex should be related to the local dynamics of the central volcanic systems (see Berrocoso et al., 2010). Furthermore, we also observe a differential behaviour between the western TN03 station and central group IZAN/TN02 stations that could be associated with a strong rheological contrast present in the centre of the island. In fact, gravity studies carried out by Gottsmann et al. (2008) detected areas with low density situated in the Northwest and high densities in the central part of the CTPV complex. These differences may affect the magnitude and nature of the deformation observed due to the distribution of the strains in the island.

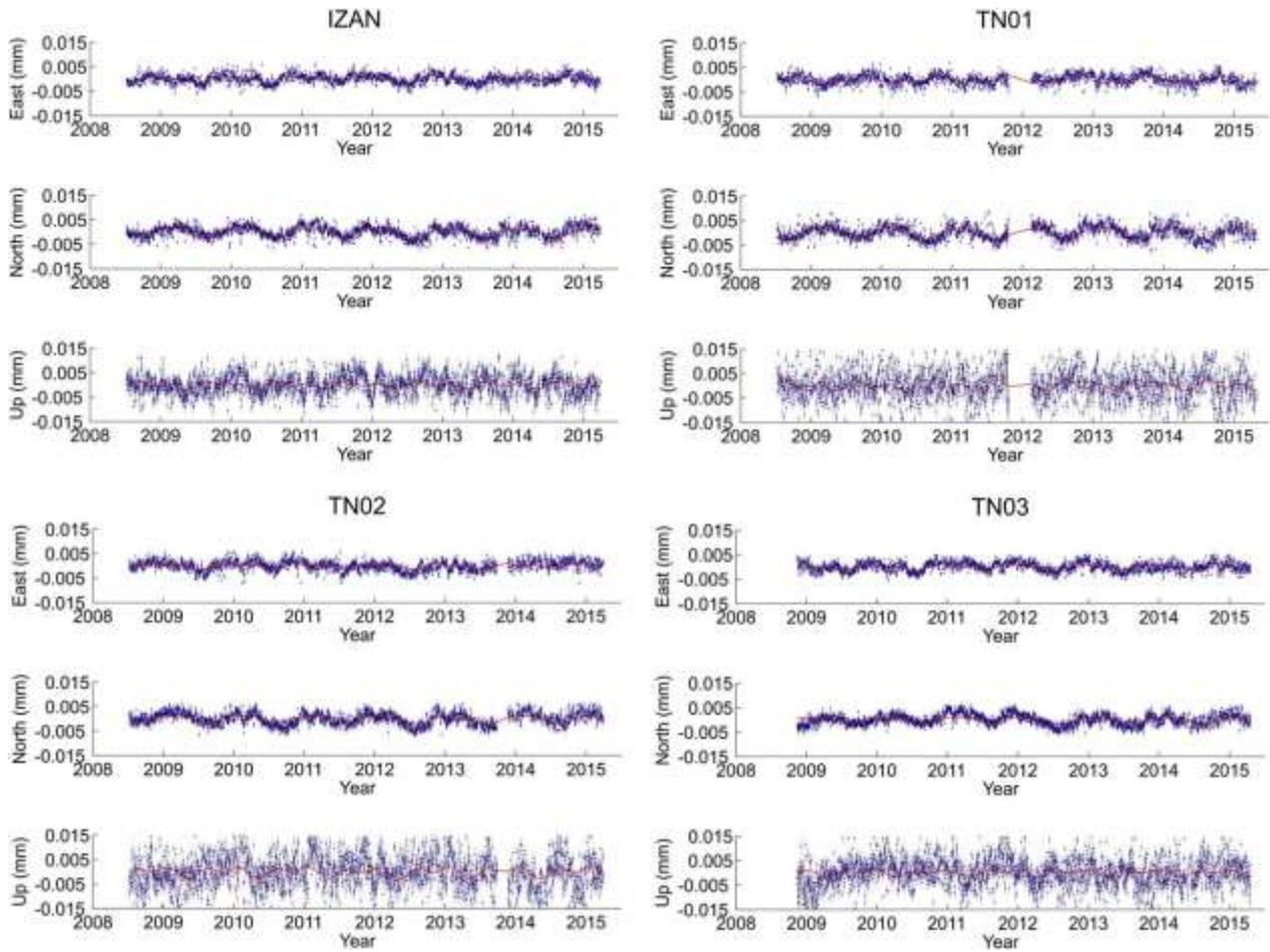

Fig. 4. De-trended position time series of the ERGNSS CGPS stations considered with error bars. East, North and Up components shown in meters. Best adjustment with annual and semiannual periodicities is also indicated.

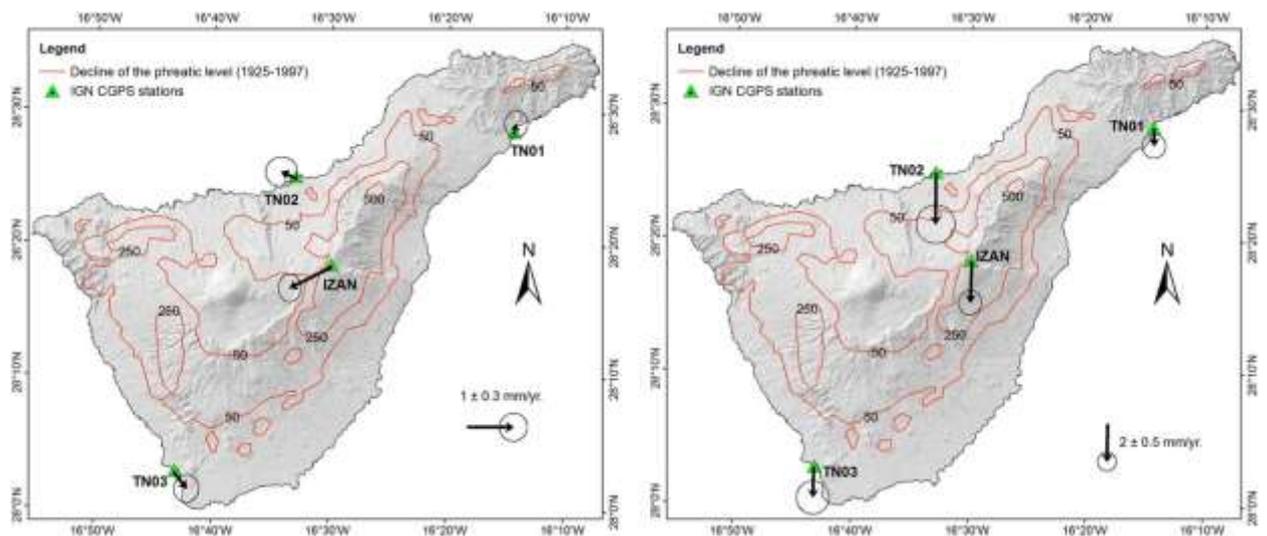

Fig. 5. Left: residual horizontal velocity field with respect to ITRF2008 plate model (Altamimi et al., 2011) considering fixed Nubia Plate in mm/yr. Right: absolute vertical velocity field in mm/yr. Decline of the phreatic level in meters from Custodio et al. (2016).

On the other hand, the horizontal residual velocities obtained are smaller than those computed by Fernández et al. (2009) in the period 2000-2006, including the 2004 sismo-volcanic crisis. This reinforces the fact that a drop of the horizontal deformation in the subsequent period (2008-2015) was detected after the 2004 unrest. We also detect a clear relation between the decline of the phreatic level of the Island (1925-1997) and a greater horizontal deformation in the IZAN CGPS station (Fig. 5). This correspondence is clearly present in IZAN and agrees with a drop of 250 m of the phreatic level under this station (Custodio et al., 2016). The direction of the residual vector is towards the inner CTPV Complex and suggests that this station may be affected by the dynamics of the central aquifer that occupies the interior of the Cañadas caldera.

The vertical velocity field in TN02, TN03 and IZAN stations show subsidence of 1.6–2.6 ± 0.4 mm/yr. during the period (2008-2015) (Table 1). Our vertical model is in line with values detected with InSAR in the CTPV complex (Fernández et al., 2009; Tizzani et al., 2010) but includes more sinking on the North and South coasts (TN02 and TN03 stations). We also note that the vertical deformation is not directly related with the decline of the phreatic level (Fig. 5). The TN02 and TN03 CGPS stations are located in areas without a significant drop of the phreatic level due to a local groundwater extraction. We suggest that the contraction of the ground under these stations may be related to a larger scale feature such as a lateral spreading of the whole volcanic edifice and the elastic lithosphere in this area (see Watts et al., 1997; Walter, 2003). On the other hand, IZAN shows a 250 meter phreatic level drop (Custodio et al., 2016). We agree with Fernández et al. (2009) that this pattern would be due partially to the contraction of the ground above the depleted aquifer, as it is also explained for the horizontal deformation observed in that station.

| Site | Absolute (mm/yr) | | | | | | Residual (mm/yr) | |
|---|---|---|---|---|---|---|---|---|
| | VE | σE | VN | σN | VU | σU | $V_rE$ | $V_rN$ |
| TN01 | 16.25 | ±0.10 | 17.32 | ±0.12 | −0.79 | ±0.26 | 0.05 | 0.20 |
| TN02 | 15.84 | ±0.14 | 17.23 | ±0.13 | −2.62 | ±0.42 | −0.34 | 0.15 |
| TN03 | 16.54 | ±0.11 | 16.73 | ±0.12 | −1.55 | ±0.36 | 0.27 | −0.33 |
| IZAN | 15.35 | ±0.10 | 16.64 | ±0.12 | −2.09 | ±0.26 | −0.88 | −0.45 |

Table 1
Absolute and residual velocities (East, North and Up) in mm/yr. Residual velocities computed using Nubian Plate as reference.

The differences observed between the stations might also suggest the existence of some rheological heterogeneities on the island that might condition the behaviour of the whole complex. In fact, the presence of high density formations with more mass under the CTPV complex and low density in other areas of the island (Gottsmann et al., 2008) could imply the different deformation patterns observed. Two more potential causes of the deformation detected and of the differences between stations could be either a local gravitational sinking of the caldera due to the weight of the Teide-Pico Viejo strato-volcano (Fernández et al., 2009) or the movement of the Azulejos graben located in Las Cañadas Caldera and currently active (Galindo et al., 2005), which would currently imply a NW-SE extension.

Although we do not discard the potential contribution of these last features on the deformation currently affecting the island of Tenerife, we consider that this is mostly due to the combination of local and a more regional causes, the significant water table drop, particular for what concerns the Cañadas aquifer, and the lateral spreading and lithospheric flexure of Tenerife, respectively. Deformations caused by subsidence of collapsed blocks in caldera structures tend to be restricted to the interior of the caldera and do not show a significant influence beyond their limits (e.g.: Battaglia et al., 2003; Folch and Gottsmann, 2006). This behaviour would not account for the deformation pattern observed here. Similarly, it is neither compatible with the potential deformation that could be associated with the movement of the Azulejos graben as this is essentially an NW-SE extension restricted to the interior of the graben located quite far from the TN01, TN02, and TN03 stations.

### 3.5 Strain rates

Based on the strain calculations, we observe a horizontal extension of 50 nstrain/yr. with an orientation of 20° around the Anaga massif (Fig. 6). This strain is orthogonal to the CTPV complex and to the extension of the Azulejos graben, and coincides with the NE ridge of the island. Furthermore, the vertical velocity model does not show marked subsidence in this area (TN01 station). This scenario would be related to the tension derived from the sinking central part applied over the more stable Anaga zone. The area around the CTPV complex presents a predominance of 70 nstrain/yr. E-W compression. This direction is orthogonal with the seismicity observed in the 2004 unrest (Almendros et al., 2007). We consider that this compression might be an expression of the subsidence detected in the area. The existence of these different zones of extension and compression marks the differences in the tectonic behaviour of the different parts of the island affected by different major structural features (caldera, rifts, …), but understanding their precise dynamics is beyond the scope of this paper. Nevertheless, the observed strain variations inform on the current state of deformation of Tenerife and on which zones are more favorable for hosting future seismovolcanic crises. In this sense, the pattern of horizontal compression and vertical subsidence would not support a continuation of the magma intrusion in the system in the period studied, which is in agreement with the low seismicity of the island detected in the period of study 2008-2015 (Domínguez Cerdeña et al., 2011; Almendros et al., 2007; www.ign.es): However, the fact that the northwestern area of the Island has low density materials (Gottsmann et al., 2008) and low P-wave velocity (García-Yeguas et al., 2012), could be related to a weak and fractured region. Additionally, this area presents some historical eruptions (Garachico (1706)) and combined with the accumulation of the E-W compression observed point to that area as the most suitable for the potential occurrence of a new volcanic crisis, in agreement with the last historical eruption (Chinyero, 1909) and the location of the seismicity associated with the 2004 event (Fig. 7).

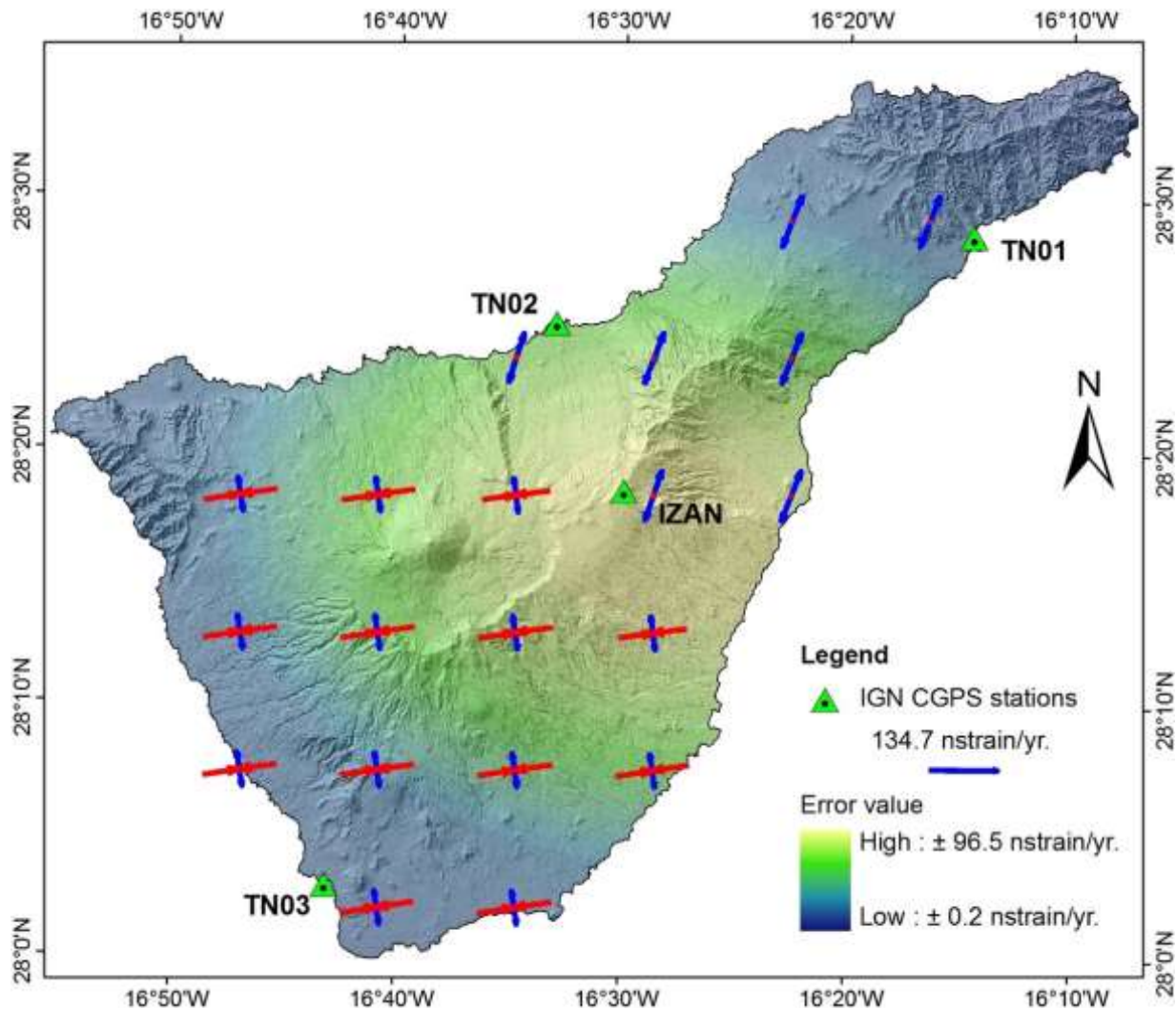

Fig. 6. Map of the principal axes of the 2D strain rate tensor in nstrain/yr. Blue and red colours show extension and contraction respectively. Magnitudes of strain errors are also included.

4. Conclusions

Using daily CGPS time series (2008–2015) processing, we have obtained a deformation model of Tenerife that reveals the existence of zones with different deformation patterns. The compression and the subsidence measured in the central-western part are consistent with a reduction of volcanic activity since the 2004 unrest (in line with the reduction of seismicity after that event). The subsidence observed can be explained by the combined effect of the lateral spreading of the elastic lithosphere affecting Tenerife and the drastic descent of the water table in the island, thus discarding other potential causes such as a local gravitational sinking of the Las Cañadas caldera, or a movement of los Azulejos graben. The extension observed on the northeastern side of the island, parallel with the trend of the Dorsal rift, is compatible with the occurrence of this central subsidence (and compression) with a more stable role of the Anaga block. Although this study provides suitable knowledge of the current deformation state of Tenerife Island, a better definition and future development of CGPS networks in the area should allow a better understanding of the Tenerife kinematics, thus improving potential models on rest-unrest dynamics of its central volcanic complex.

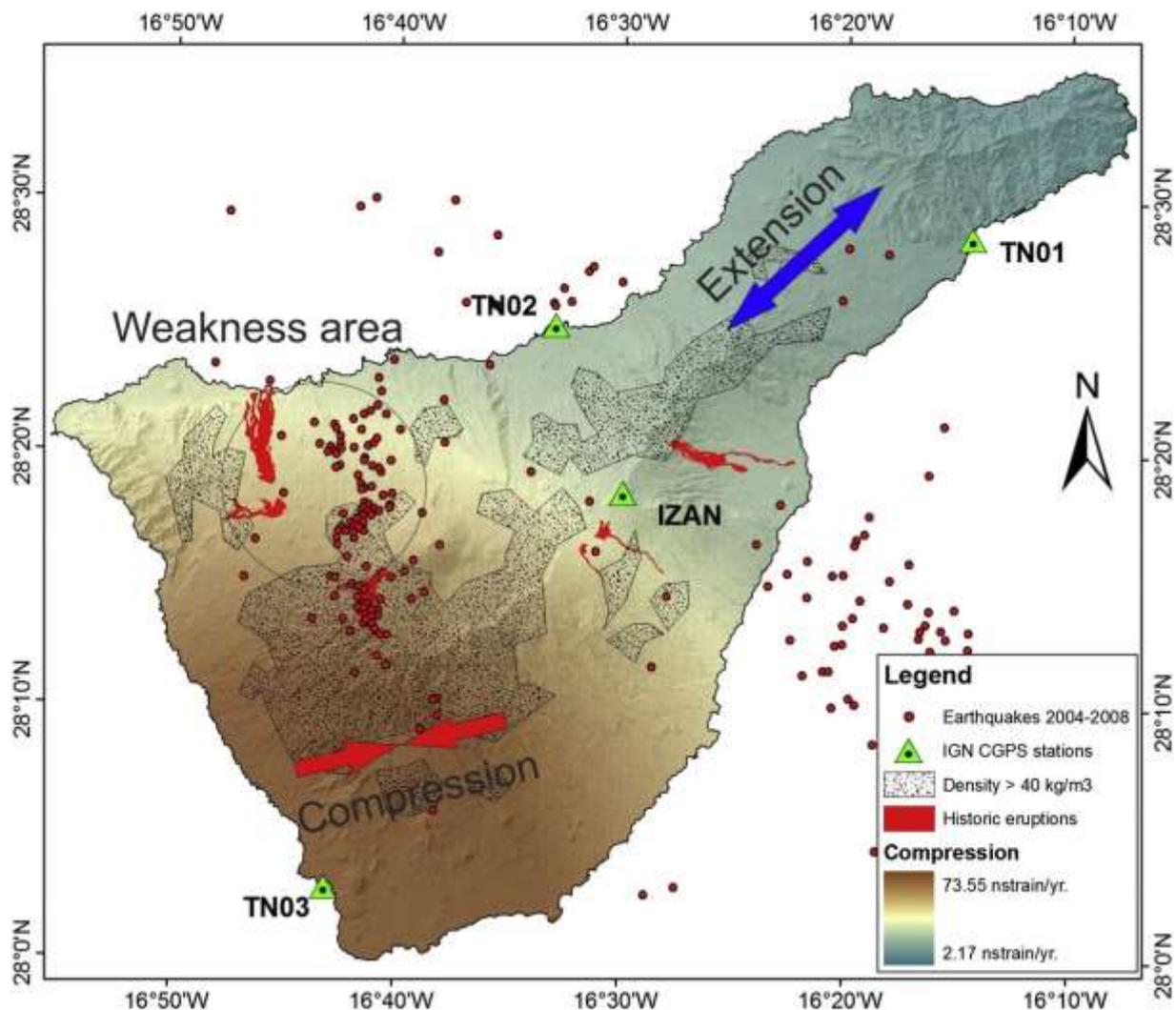

Fig. 7. Geodynamic scheme of Tenerife Island. The north-westward compression agrees with low density materials (Gottsmann et al., 2008) and low P-wave velocity (García-Yeguas et al., 2012). The presence of historic eruptions, seismicity during 2004 volcanic crisis and a measurable E-W compression, suggest that the NW area of weakness could be a source of future volcanic activity. Seismicity during the 2004 unrest (red dots) and historic eruptions (areas in red) are also shown.

Acknowledgements

The methodology described in this study is part of A. Sánchez-Alzola's PhD Thesis at Department of Cartographic, Geodetic Engineering and Photogrammetry of Jaén University. The authors thank the National Geographic Institute of Spain for the ERGNSS network GPS data and the two anonymous referees in the revision stage of the paper for their constructive comments.